\documentclass{article}
\usepackage{times}
\usepackage{amsfonts}
\begin{document}
\noindent
{\Large  THE HOLOGRAPHIC QUANTUM}
\vskip1cm
\noindent
{\bf P. Fern\'andez de C\'ordoba}$^{a}$, {\bf  J.M. Isidro}$^{b}$ and {\bf J. Vazquez Molina}$^{c}$\\
Instituto Universitario de Matem\'atica Pura y Aplicada,\\ Universidad Polit\'ecnica de Valencia, Valencia 46022, Spain\\
${}^{a}${\tt pfernandez@mat.upv.es}, ${}^{b}${\tt joissan@mat.upv.es},\\ $^{c}${\tt joavzmo@etsii.upv.es}
\vskip.5cm
\noindent
\today
\vskip.5cm
\noindent
{\bf Abstract}  We present a map of standard quantum mechanics onto a dual theory, that of the classical thermodynamics of irreversible processes. While no gravity is present in our construction, our map exhibits features that are reminiscent of the holographic principle of quantum gravity.

\tableofcontents
\vskip.5cm

\section{Introduction}\label{uno}

The holographic principle \cite{BOUSSO, THOOFT0, SUSSKIND1} has permeated wide areas of theoretical physics over the last twenty years. Stepping outside its initial quantum--gravity framework, it reached string theory \cite{MALDACENA, WITTEN} as well as more established domains such as QCD \cite{QCD} and condensed matter theory \cite{CMT}, to name but a few. 

Another theoretical development of recent years is the recognition that {\it gravity arises as an emergent phenomenon}\/ \cite{PADDY2, PADDY3, VERLINDE}, a fact that has far--reaching consequences for our understanding of spacetime. Added to the dissipative properties already known to be exhibited by gravity  \cite{THOOFT1, PENROSE1, SMOLIN1, SMOLIN2}, this opens the gate to the application of thermodynamics to (supposedly) nonthermal physics. Indeed, thermodynamics is the paradigm of emergent theories. It renounces the knowledge of a vast amount of detailed microscopic information, keeping just a handful of macroscopic variables  such as volume, pressure and temperature---sufficient to state robust physical laws of almost universal applicability. These macroscopic variables are coarse--grained averages over the more detailed description provided by some underlying, microscopic degrees of freedom. Which brings us to yet another theoretical breakthrough of recent times that is worthy of mention: the notion of {\it emergence}\/ \cite{CARROLL}.

The property of emergence has been postulated not only of gravity, but also of Newtonian mechanics \cite{VERLINDE} and of quantum mechanics \cite{ELZE, THOOFT2}; a key concept here is that of an {\it entropic force}\/. Equipped with thermodynamical tools as befits any emergent theory, we have in refs. \cite{NOI1, NOI2, NOI3} developed a framework that maps semiclassical quantum mechanics onto the classical thermodynamics of irreversible processes in the linear regime, the latter as developed by Onsager, Prigogine and collaborators \cite{ONSAGER, PRIGOGINE}. Within this framework,  the statement often found in the literature, {\it  quantisation is dissipation}\/ \cite{BLASONE}, can be given a new interpretation.

In this paper we elaborate further on the above--mentioned map of semiclassical quantum mechanics onto the classical theory of linear, irreversible processes (sections \ref{srftrm} and \ref{dos}); we call these two theories {\it dual}\/ to each other. From there we move on to the nonlinear regime of the thermodynamics or, equivalently, to the quantum regime beyond the Gaussian approximation (section \ref{dduee}). Next we formulate a holographic--like principle for quantum mechanics (section \ref{tres}) and place it in correspondence with the second law of thermodynamics (section \ref{cuatro})).    The term {\it holographic--like}\/ is meant to stress that, while it is true that no gravity is present in our framework, an undeniable conceptual similarity with the holographic principle of quantum gravity underlies the principle postulated here. We summarise our conclusions in section \ref{cinco}.

A word on notation is in order. Rather than using natural units, we will explicitly retain Planck's constant $\hbar$ and Boltzmann's constant $k_B$ in our expressions, in order to better highlight the properties of the map presented here between quantum mechanics and irreversible thermodynamics. In particular, the role that $\hbar$ plays on the mechanical side of our correspondence will be played by $k_B$ on the thermodynamical side. If we were to set $\hbar=1=k_B$, the fact that they are counterparts under our correspondence \cite{COHEN, PADDY1} would be somewhat obscured.

\section{Basics in irreversible thermodynamics}\label{srftrm}

The following is a very brief summary of some notions of irreversible thermodynamics \cite{ONSAGER, PRIGOGINE} that we will make use of.

Let an irreversible thermodynamical system be characterised by its entropy function $S$. Assume that the thermodynamical state of the system is determined by just one extensive variable $x=x(\tau)$, where $\tau$ is time variable. We can thus write $S=S(x(\tau))$. At any instant of time, the probability $P$ of a state is given by Boltzmann's principle,
\begin{equation}
k_B\ln P=S+ {\rm const}.
\label{foman}
\end{equation}
Let $S_0$ denote the maximum (equilibrium) value of $S$, and let us redefine the coordinate $x$ so it will vanish when evaluated at equilibrium: $S_0=S(x=0)$. Irreversible thermodynamics \cite{ONSAGER} analyses the response of the system when driven away from equilibrium. For this purpose one introduces the {\it thermodynamical force}\/ $X$,
\begin{equation}
X=\frac{{\rm d}S}{{\rm d}x},
\label{stark}
\end{equation}
which measures the tendency of the system to restore equilibrium. Nonequilibrium causes fluxes to appear in the system, that is, nonvanishing 
time derivatives ${\rm d}x/{\rm d}\tau$ and ${\rm d}S/{\rm d}\tau$. Further one supposes that the irreversible process considered is {\it linear}\/. This amounts to the assumption that the flux is proportional to the force,
\begin{equation}
\frac{{\rm d}x}{{\rm d}\tau}=LX, \qquad L>0,
\label{lineale}
\end{equation}
where $L$ is a positive constant, independent of $x$ and $\tau$.  One also writes (\ref{lineale}) under the form
\begin{equation}
X=R\frac{{\rm d}x}{{\rm d}\tau}, \qquad R=L^{-1}>0,
\label{premi}
\end{equation}
where the dimensions of $R$ are time $\times$ entropy $\times\; x^{-2}$. Eq. (\ref{premi}) is often termed a {\it phenomenological law}\/. Indeed numerous dissipative phenomena, at least to first order of approximation, take on the form of a linear relation between a driving force $X$ and the corresponding flux ${\rm d}x/{\rm d}\tau$: Ohm's law in electricity, Fourier's law of heat transfer, etc, are familiar examples. In linear irreversible thermodynamics, the time rate of entropy production is the product of those two:
\begin{equation}
\frac{{\rm d}S}{{\rm d}\tau}=X\frac{{\rm d}x}{{\rm d}\tau}.
\label{silenz}
\end{equation}

On the other hand, Taylor--expanding the entropy around its (maximum) equilibrium value and keeping terms up to second--order we have
\begin{equation}
S=S_0-\frac{1}{2}sx^2+\ldots, \qquad s:=-\left(\frac{{\rm d}^2S}{{\rm d}x^2}\right)_0>0.
\label{teilo}
\end{equation}
Three consequences follow from truncating the expansion (\ref{teilo}) at second order. First, the force $X$ is a linear function of the coordinate $x$:
\begin{equation}
X=-sx.
\label{llin}
\end{equation}
Second, in conjunction with Boltzmann's principle (\ref{foman}), the expansion (\ref{teilo}) implies that the probability distribution for fluctuations is a Gaussian in the extensive variable $x$:
\begin{equation}
P(x)=Z^{-1}\exp\left(\frac{S}{k_B}\right)=Z^{-1}\exp\left(-\frac{1}{2k_B}sx^2\right),
\label{sarto}
\end{equation}
where $Z$ is some normalisation.\footnote{We will henceforth omit all normalistion factors, bearing in mind that all probabilites are to be normalised at the end.} Third, the phenomenological law (\ref{premi}) specifies a linear submanifold of thermodynamical phase space:
\begin{equation}
R\frac{{\rm d}x}{{\rm d}\tau}+sx=0.
\label{gus}
\end{equation}

Fluctuations around the deterministic law given by Eq. (\ref{gus}) can be modelled by the addition of a random force $F_{r}$. This turns the deterministic equation (\ref{gus}) into the stochastic equation
\begin{equation}
R\frac{{\rm d}x}{{\rm d}\tau}+sx=F_r.
\label{unica}
\end{equation}
We are interested in computing the path $x=x(\tau)$  under the influence of these random forces, under the assumption that $F_r$ has a vanishing average value. While mimicking random fluctuations, this assumption ensures that the net force continues to be given as in the deterministic Eq. (\ref{gus}). Now our aim is to calculate the probability of any path in configuration space. For this purpose we need to introduce some concepts borrowed from ref. \cite{DOOB}. 

The {\it unconditional probability density function}\/  $f\left({x\atop\tau}\right)$, also called {\it one--gate function}\/, is defined such that the product
$f\left({x\atop\tau}\right){\rm d}x$ equals the probability that the random trajectory $x=x(\tau)$ pass through a gate of width ${\rm d}x$ around $x$ at the instant $\tau$. The {\it conditional probability density function}\/ $f\left({x_2\atop \tau_2}{\Big \vert}{x_{1}\atop \tau_{1}}\right)$, also called the {\it two--gate function}\/, is defined such that $f\left({x_2\atop \tau_2}{\Big \vert}{x_{1}\atop \tau_{1}}\right){\rm d}x_2\,{\rm d}x_{1}$ equals the probability that a thermodynamical path pass through a gate of width ${\rm d}x_2$ around $x_2$ at time $\tau_2$, {\it given}\/ that it passed through a gate of width ${\rm d}x_{1}$ around $x_1$ at time $\tau_{1}$. The assumption that our stochastic process (\ref{unica}) satisfies the Markov property ensures that the unconditional probability $f\left({x_2\atop\tau_2}\right)$ can be obtained from the conditional probability $f\left({x_2\atop\tau_2}{\Big\vert}{x_1\atop\tau_1}\right)$  by letting $\tau_1=-\infty$ in the latter and setting a fixed value of $x_1$, say $x_1=0$. Informally speaking: Markov systems have a short--lived memory.

Let us consider a time interval $(\tau_1,\tau_{n+1})$ , which we divide into $n$ subintervals of equal length. Then the conditional probabilities obey the Chapman--Kolmogorov equation,
\begin{equation}
f\left({x_{n+1}\atop \tau_{n+1}}{\Big\vert}{x_1\atop \tau_1}\right)=\int{\rm d}x_n\cdots\int{\rm d}x_2\, f\left({x_{n+1}\atop \tau_{n+1}}{\Big\vert}{x_n\atop \tau_n}\right)\cdots f\left({x_{2}\atop \tau_{2}}{\Big\vert}{x_1\atop \tau_1}\right),
\label{muul}
\end{equation}
where all $n-1$ intermediate gates at $x_2, x_3,\ldots, x_n$ are integrated over.  In particular, the unconditional probability density $f\left({x\atop \tau}\right)$ propagates according to the law
\begin{equation}
f\left(x_2\atop \tau_2\right)=\int{\rm d}x_1\,f\left({x_2\atop\tau_2}{\Big\vert}{x_1\atop\tau_1}\right)f\left(x_1\atop \tau_1\right).
\label{despi}
\end{equation}
It turns out that, for a Markovian Gaussian process, the conditional probability function $f\left({x_{2}\atop\tau_2}{\Big\vert}{x_{1}\atop \tau_1}\right)$ that solves the Chapman--Kolmogorov equation is given by \cite{ONSAGER}
\begin{equation}
f\left({x_{2}\atop\tau_2}{\Big\vert}{x_{1}\atop \tau_1}\right)=\frac{s}{2k_B}\frac{{\rm e}^{s(\tau_2-\tau_1)/2R}}
{\sqrt{\pi\,\sinh\left[s(\tau_2-\tau_1)/R\right]}}
\label{prpazz}
\end{equation}
$$
\times
\exp\left\{-\frac{s}{2k_B}\frac{\left[{\rm e}^{s(\tau_2-\tau_1)/2R}\,x_{2}-{\rm e}^{-s(\tau_2-\tau_1)/2R}\,x_{1}\right]^2}{2\,\sinh\left[s(\tau_2-\tau_1)/R\right]}\right\}.
$$
As a consistency check we observe that, in the limit $\tau_2\to\infty$, the conditional probability (\ref{prpazz}) reduces to the unconditional probability (\ref{sarto}). Using the Chapman--Kolmogorov equation (\ref{muul}) one can reexpress the conditional probability (\ref{prpazz}) as 
\begin{equation}
f\left({x_{n+1}\atop \tau_{n+1}}{\Big\vert}{x_1\atop \tau_1}\right)=\exp\left[-\frac{1}{4k_B}\int_{\tau_1}^{\tau_{n+1}}{\rm d}\tau\,R\left(\frac{{\rm d}x}{{\rm d}\tau}+\gamma x\right)^2\right]_{\rm min},  \gamma:=\frac{s}{R},
\label{kondbrop}
\end{equation}
subject to the boundary conditions $x(\tau_1)=x_1$ and $x(\tau_{n+1})=x_{n+1}$. Above, $\gamma$ carries the dimension of inverse time, while the subscript {\it min}\/ reminds us that the integral is to be evaluated along that particular path which minimises the integral.

Now $f\left({x_2\atop\tau_2}\right)$ can be obtained from $f\left({x_2\atop\tau_2}{\Big\vert}{x_1\atop\tau_1}\right)$  by letting $\tau_1=-\infty$ and $x_1=0$ in the latter. In order to take this limit in Eq. (\ref{kondbrop}) we first define the {\it thermodynamical Lagrangian}\/ ${\cal S}$ to be
\begin{equation}
{\cal S}:=\frac{R}{2}\left(\frac{{\rm d}x}{{\rm d}\tau}+\gamma x\right)^2,
\label{colaborr}
\end{equation}
or, dropping a total derivative,
\begin{equation}
{\cal S}=\frac{R}{2}\left[\left(\frac{{\rm d}x}{{\rm d}\tau}\right)^2+\gamma^2x^2\right].
\label{muntagra}
\end{equation}
The dimensions of ${\cal S}$ are entropy per unit time. The corresponding Euler--Lagrange equation reads
\begin{equation}
\frac{{\rm d}^2x}{{\rm d}\tau^2}-\gamma^2x=0, 
\label{tomo}
\end{equation}
while 
\begin{equation}
x(\tau)=x_2{\rm e}^{\gamma(\tau-\tau_2)}
\label{disi}
\end{equation}
is the particular solution to (\ref{tomo}) that satisfies the boundary conditions $x(\tau=-\infty)=0$ and $x(\tau=\tau_2)=x_2$. Thus evaluating (\ref{kondbrop}) along this extremal path yields
\begin{equation}
f\left({x_2\atop\tau_2}{\Big\vert}{0\atop -\infty}\right)=f\left(x_2\atop \tau_2\right)=\exp\left[-\frac{s}{2k_B}(x_2)^2\right].
\label{konss}
\end{equation}
This is again in agreement with Boltzmann's principle (\ref{foman}) in the Gaussian approximation (\ref{teilo}).  
Moreover, the conditional probability density $f\left({x_2\atop\tau_2}{\Big\vert}{x_1\atop\tau_1}\right)$ admits the path--integral representation \cite{ONSAGER}\footnote{What quantum theorists call the Feynman path integral was independently developed in ref. \cite{ONSAGER} by Onsager and collaborators, who appear to have arrived at the notion of a path integral all by themselves, without previous  knowledge of Feynman's earlier work \cite{FEYNMAN}.} 
\begin{equation}
f\left({x_{2}\atop \tau_{2}}{\Big\vert}{x_1\atop \tau_1}\right)=\int_{x(\tau_1)=x_1}^{x(\tau_2)=x_2}{\rm D}x(\tau)\,
\exp\left\{-\frac{1}{2k_B}\int_{\tau_1}^{\tau_{2}}{\rm d}\tau\,{\cal S}\right\}.
\label{cecece}
\end{equation}
In fact, a saddle--point evaluation of the path integral (\ref{cecece}) is readily seen to yield the two--gate function (\ref{kondbrop}). 

The above Eqs. (\ref{stark})--(\ref{cecece}) have obvious generalisations to a case with $D$ independent thermodynamical coordinates.

\section{Quantum mechanics vs. irreversible thermodynamics}\label{dos}

The attentive reader will have noticed the striking similarity between Eqs. (\ref{stark})--(\ref{cecece}) and the quantum mechanics of the harmonic oscillator. The corresponding Lagrangian is
\begin{equation}
{\cal L}=\frac{m}{2}\left(\frac{{\rm d}x}{{\rm d}t}\right)^2-\frac{k}{2}x^2.
\label{giannna}
\end{equation}
Mechanical time is denoted by the variable $t$; it is related to thermodynamical time $\tau$ through the Wick rotation
\begin{equation}
\tau={\rm i}t.
\label{wieck}
\end{equation}
We define as usual the angular frequency $\omega$ through $\omega^2=k/m$. Let us for simplicity assume that the thermodynamical extensive coordinate $x$ of the dual irreversible thermodynamics is a length. In this way no dimensionful factor is needed to reinterpret it as the coordinate of the harmonic oscillator in the mechanical dual theory. Then the Wick rotation (\ref{wieck}) and the replacements\footnote{Implicit in the replacements (\ref{diktionaer}) is the assumption that the thermodynamical extensive variable $x$, and the mechanical variable $x$, both have units of length. A dimensionful conversion factor is to be understood in case the dimensions do not match.}
\begin{equation}
\frac{m\omega}{\hbar}=\frac{s}{2k_B}, \qquad \omega=\gamma
\label{diktionaer}
\end{equation}
provide us with a dictionary to establish a 1--to--1 map between the linear, irreversible thermodynamics of section \ref{srftrm} and the quantum mechanics of the harmonic oscillator. 

Specifically, let us spell out the entries of this map, one by one \cite{EMERHENTE}. The mechanical Lagrangian (\ref{giannna}) is readily obtained from its thermodynamical counterpart (\ref{muntagra}) upon application of the replacements (\ref{wieck}), (\ref{diktionaer}): 
\begin{equation}
\frac{{\cal S}}{2k_B}=-\frac{{\cal L}}{\hbar}.
\label{rotuik}
\end{equation}
The above also makes it clear that the thermodynamical analogue of Planck's constant $\hbar$ is twice Boltzmann's constant, $2k_B$. In this way
the thermodynamical path integral (\ref{cecece}) becomes its usual quantum--mechanical expression. Unconditional probabilities $f\left({x\atop\tau}\right)$ in thermodynamics become wavefunctions squared $\vert\psi(x,t)\vert^2$ in quantum mechanics. Thus the 1--gate distribution function (\ref{konss}) gives the squared modulus of the oscillator groundstate,
\begin{equation}
f\left({x\atop {\rm i}t}\right)=\exp\left(-\frac{m\omega}{\hbar}x^2\right).
\label{karch}
\end{equation}
The thermodynamical conditional probabiliy (\ref{prpazz}) becomes proportional to the quantum--mechanical Feynman propagator. Away from the caustics, the latter is given by
\begin{equation}
K\left(x_2,t_2\vert x_1,t_1\right)=\sqrt{\frac{m\omega}{2\pi{\rm i}\hbar\sin\left(\omega(t_2-t_1)\right)}}
\label{kaus}
\end{equation}
$$
\times \exp\left\{\frac{{\rm i}m}{2\hbar}\frac{\omega}{\sin\left(\omega(t_2-t_1)\right)}\left[(x_2^2+x_1^2)\cos\left(\omega(t_2-t_1)\right)-
2x_2x_1\right]\right\}
$$
and one actually finds
\begin{equation}
f\left({x_2\atop {\rm i}t}{\Big\vert}{x_1\atop 0}\right)=\exp\left(\frac{{\rm i}\omega t}{2}-\frac{\Delta V}{\hbar\omega}\right)\sqrt{\frac{2m\omega}{\hbar}}\,K\left(x_2,t\vert x_1,0\right),
\label{novita}
\end{equation}
where $\Delta V=V(x_2)-V(x_1)$, with $V(x)=kx^2/2$ the harmonic potential. The Chapman--Kolmogorov equation (\ref{muul}) becomes the group property of propagators, while the propagation law (\ref{despi}) exactly matches that for wavefunctions $\psi$ under propagators $K$. Altogether, the promised 1--to--1 map is complete.

Our Eqs. (\ref{giannna})--(\ref{novita}) have obvious generalisations to higher dimensions. Since the concept of {\it equipotential submanifolds}\/ will play a key role in our duality between quantum mechanics and irreversible thermodynamics, it will be useful to consider the lowest dimension in which equipotential manifolds are {\it 2--dimensional surfaces}\/. Configuration space is then 3--dimensional, which we take to be $\mathbb{R}^3$, coordinatised by $x,y,z$. For simplicity we will assume the harmonic potential to be isotropic, so the harmonic force is ${\bf F}_h=-k(x,y,z)$. On the thermodynamical side of our correspondence, this translates into the fact that Onsager's (inverse) coefficients $R_x$, $R_y$, $R_z$  in Eq. (\ref{premi}) are all equal, so the dissipative force acting on the system is ${\bf F}_d=R({\rm d}x/{\rm d}\tau, {\rm d}y/{\rm d}\tau, {\rm d}z/{\rm d}\tau)$.
We then have a thermodynamical Lagrangian
\begin{equation}
{\cal S}=\frac{R}{2}\left[\left(\frac{{\rm d}x}{{\rm d}\tau}\right)^2+\left(\frac{{\rm d}y}{{\rm d}\tau}\right)^2+\left(\frac{{\rm d}z}{{\rm d}\tau}\right)^2+\gamma^2(x^2+y^2+z^2)\right]
\label{thermlangg}
\end{equation}
and a mechanical Lagrangian
\begin{equation}
{\cal L}=\frac{m}{2}\left[\left(\frac{{\rm d}x}{{\rm d}t}\right)^2+\left(\frac{{\rm d}y}{{\rm d}t}\right)^2+\left(\frac{{\rm d}z}{{\rm d}t}\right)^2-\omega^2(x^2+y^2+z^2)\right].
\label{bahaaltta}
\end{equation}
The latter has the family of 2--dimensional spheres $x^2+y^2+z^2=\rho^2$ as equipotential surfaces within the mechanical configuration space $\mathbb{R}^3$.  We claim that the thermodynamical counterpart of this family of spheres is the family of 5--dimensional submanifolds
\begin{equation}
\left(\frac{{\rm d}x}{{\rm d}\tau}\right)^2+\left(\frac{{\rm d}y}{{\rm d}\tau}\right)^2+\left(\frac{{\rm d}z}{{\rm d}\tau}\right)^2+\gamma^2(x^2+y^2+z^2)=\rho^2
\label{suisse}
\end{equation}
within the thermodynamical phase space $\mathbb{R}^6$; we may call the above hypersurfaces {\it isoentropic submanifolds}\/. Although we seem to have a dimensional mismatch between isoentropic submanifolds and equipotential surfaces, this mismatch disappears if we restrict to those thermodynamical trajectories that satisfy the equation of motion of the thermodynamical Lagrangian (\ref{thermlangg}). This equation was given in (\ref{tomo}) and solved in (\ref{disi}); we see that, {\it on shell}\/, the velocity ${\rm d}x/{\rm d}\tau$ is proportional to the coordinate $x$. This property effectively allows us to replace the term $({\rm d}x/{\rm d}\tau)^2+({\rm d}y/{\rm d}\tau)^2+({\rm d}z/{\rm d}\tau)^2$ in Eq. (\ref{suisse}) with a constant multiple of $x^2+y^2+z^2$. In turn, this reduces the family of 5--dimensional submanifolds (\ref{suisse}) to a family of 2--dimensional spheres---exactly as in the mechanical case.

We conclude that {\it equipotential surfaces for the mechanical problem become isoentropic surfaces for the thermodynamical problem, and viceversa}\/. This is in nice agreement with the results of ref. \cite{VERLINDE} for the gravitational potential, in the context of a theory of emergent spacetime.

\section{Beyond the harmonic approximation}\label{dduee}

While explicit expressions for our map between quantum mechanics and irreversible thermodynamics are difficult to obtain beyond the harmonic approximation considered so far, some key physical ideas can be extracted from the previous analysis and generalised to an arbitrary potential. On the thermodynamical side, this generalisation implies going beyond the Gaussian approximation made in Eq. (\ref{teilo}) or, equivalently, beyond the assumption (\ref{llin}) of linearity between forces and fluxes.

Let a mechanical system be described by a Lagrangian function ${\cal L}={\cal L}(q_i,\dot q_i)$.  For simplicity we assume our configuration space to be $\mathbb{R}^D$; an additional $\mathbb{R}$ stands for the time axis. The mechanical time variable $t$, initially real, will be complexified presently. 

We will equate certain spacetime concepts (on the left--hand side of the equations below) to certain thermodynamical quantities (on the right). To begin with, we observe that the two physical constants $\hbar$ and $k_B$ allow one to regard time $t$ and temperature $T$ as mutually inverse, through the combination
\begin{equation}
\frac{1}{t}=\frac{k_B}{\hbar}T.
\label{sumario}
\end{equation}
Admittedly, this observation is not new \cite{DEBROGLIE}.

Corresponding to the mechanical system governed by the Lagrangian ${\cal L}(q_i,\dot q_i)$ there will be a thermodynamical system whose dynamics will be governed by an entropy $S=\int{\cal S}{\rm d}t$. Following our previous result (\ref{rotuik}), let us postulate the following differential relation between the two of them:
\begin{equation}
\frac{1}{\hbar}{\cal L}{\rm d}t=\frac{C}{2k_B}{\rm d}S=\frac{C}{2k_B}{\cal S}{\rm d}t,\qquad C\in\mathbb{C}.
\label{refinatissimo}
\end{equation}
Again, dimensionality arguments basically fix the two sides of the above relation, but leave room for a dimensionless number $C$. Agreement with the  Wick rotation (\ref{wieck}) requires that we set $C=-{\rm i}$.  Now Eq. (\ref{refinatissimo}) overlooks the fact that the right--hand side  contains the exact differential ${\rm d}S$, while the differential ${\cal L}{\rm d}t$ on the left--hand side is generally {\it not}\/ exact. In other words, while there exists a well--defined entropy {\it function}\/ $S=\int{\cal S}{\rm d}t$, the line integral $I=\int {\cal L}{\rm d}t$ generally depends on the trajectory in $\mathbb{R}^D$ being integrated along.

The mechanical action $I$, however, {\it can}\/ define a path--independent function of the integration endpoint if we restrict to a certain class of trajectories in $\mathbb{R}^D$. Let us see how this comes about. Let $V=V(q_i)$ be the potential function of the mechanical system under consideration. The equation
\begin{equation}
V(q_i)={\rm const}
\label{equi}
\end{equation}
defines, as the constant on the right--hand side is varied, a family of  $(D-1)$--dimensional, {\it equipotential submanifolds}\/ of $\mathbb{R}^D$. An elementary example, when $D=3$, is the case of the Newtonian potential generated by a point mass located at the origin $O$. Then the above family of equipotential surfaces is a family of concentric spheres $\mathbb{S}_\rho$ of increasing radii $\rho>0$, all centred at $O$. This family of equipotentials, singular only at $O$, defines a foliation of $\mathbb{R}^3-\{O\}$, so the latter space equals the union $\cup_{\rho>0}\mathbb{S}_\rho$ of all leaves $\mathbb{S}_\rho$. This foliation can also be used to define a coordinate system on $\mathbb{R}^3-\{O\}$. Namely, one splits $\mathbb{R}^3-\{O\}$ into 2 tangential directions to the spheres of the foliation, and 1 normal direction. For example, the standard spherical coordinates $\rho, \theta, \varphi$ centred at $O$ qualify as such a coordinate system, $\rho$ being the normal coordinate and $\theta, \varphi$ the tangential coordinates.

Returning now to the general case when both $D$ and $V(q_i)$ are arbitrary, Eq. (\ref{equi}) defines, for each particular value of the constant on the right--hand side, one equipotential leaf $\mathbb{L}_n$ of a foliation $\cup_{n}\mathbb{L}_n$ of $\mathbb{R}^D$. Here the subindex $n$ stands for a certain (local) coordinate $n$ on $\mathbb{R}^D$ that is normal to all the leaves. The $D-1$ tangential coordinates thus span the $(D-1)$--dimensional leaves $\mathbb{L}_n$, each one of them being located at a specific value of the normal coordinate $n$. We will assume that all the leaves $\mathbb{L}_n$ are compact. 

Trajectories within $\mathbb{R}^D$ that run exclusively along this normal coordinate $n$, thus being orthogonal to the leaves, are such that the action integral $I$ {\it does}\/ defines a function $I_n$ of the integration endpoint; the subindex $n$ reminds us of the restriction to these normal trajectories. Independence of path is merely a consequence of the 1--dimensionality of the normal directions to the equipotential leaves $\mathbb{L}_n$. This is the particular class of trajectories mentioned above: along them, ${\cal L}{\rm d}t$ defines an exact differential, ${\rm d}I_{n}$. For these normal trajectories, the differential equation (\ref{refinatissimo}) makes perfect sense as an equality between two exact differentials. For these normal trajectories we can write
\begin{equation}
\frac{1}{\hbar}I_{n}-\frac{C}{2k_B}S={\rm const}.
\label{raffinato}
\end{equation}

Now the sought--for thermodynamics {\it cannot}\/ be the standard thermodynamics of equilibrium processes as presented in any standard textbook, say, ref. \cite{CALLEN}. Among other reasons for this not being the case, standard equilibrium thermodynamics does not include time as one of its variables.  We have already in section \ref{dos} produced evidence that it must in fact be the {\it explicitly time--dependent,  classical thermodynamics of irreversible processes}\/ as developed by Onsager, Prigogine {\it et al}\/ \cite{ONSAGER, PRIGOGINE}. We will present arguments in section \ref{tres}, to the effect that quantum states arise through a dissipative mechanism. For completeness the thermodynamical dual to quantum mechanics must be supplemented with the relation
\begin{equation}
\frac{1}{T}=\frac{\partial S}{\partial U}, 
\label{grund}
\end{equation}
which must always be satisfied. So we take  (\ref{grund}) to {\it define}\/ the internal energy $U$ of the thermodynamical theory, given that $T$ and $S$ have already been defined.

\section{Quantum states as equivalence classes of classical trajectories}\label{tres}

A key consequence of using normal and tangential coordinates in $\mathbb{R}^D$ is that quantum states $\psi$, to be constructed presently, will factorise as
\begin{equation}
\psi=\psi_t\psi_n,
\label{fakto}
\end{equation}
or sums thereof. Here, the normal wavefunction $\psi_n$ depends exclusively on the normal coordinate $n$, while $\psi_t$ is a function of the tangential coordinates. For example, in the case of the Coulomb potential, the wavefunction $\psi_t$ would be a spherical harmonic $Y_{lm}(\theta, \varphi)$, while $\psi_n$ would be a radial wavefunction $R_{nl}(\rho)$. This construction contains elements that are very reminiscent of those present in ref. \cite{VERLINDE}. In this latter paper, {\it equipotential surfaces of the gravitational potential are identified as isoentropic surfaces}\/. Our equipotential leaves are the counterpart of the {\it holographic screens}\/ of ref. \cite{VERLINDE}. 

Moreover, the classical mechanics exhibits a precise mechanism whereby {\it different classical trajectories coalesce into a single equivalence class that can, following ref. \cite{THOOFT2}, be identified as a single quantum state $\psi$}\/. So the presence of Planck's constant $\hbar$ in Eq. (\ref{refinatissimo}) obeys not just dimensional reasons---it is the sure sign of an information--loss mechanism, a dissipative processs that is truly quantum in nature. 

Let us see how this dissipation comes about. In order to do this we need to explain why many different classical trajectories coalesce into one single quantum state $\psi$.  A quantum of area on the leaf $\mathbb{L}_n$ measures $L_P^2$, where $L_P$ denotes the Planck length. According to the holographic principle, at most 1 bit of information fits into this quantum of area $L_P^2$. One classical trajectory traversing this quantum of area corresponds to 1 bit of information. Classically one can regard the surface density of trajectories as being correctly described by a smooth distribution function: there fit some $1.4\times 10^{69}$ classical trajectories into each square meter of area on the leaf $\mathbb{L}_n$\cite{BOUSSO}. Although this is a huge number, it sets an upper limit on the potentially infinite number of classical trajectories that can traverse one quantum of area $L_P^2$.

The holographic principle alone would suffice to account for the lumping together of many different classical trajectories into one equivalence class. One equivalence class, or quantum state, would be comprised by all those different classical trajectories crossing one given quantum of area $L_P^2$. 

Of course, the {\it actual}\/ number of quantum particles traversing one square meter of area on the leaf $\mathbb{L}_n$ is much smaller than the above $1.4\times 10^{69}$. The reason is simple: quantum effects become nonnegligible on matter well before quantum--gravity effects become appreciable on the geometry. Again, the existence of a (now particle--dependent) quantum of area is responsible for this. This can be seen as follows. 

Let $m$ be the mass of the particle under consideration. Its Compton wavelength $\lambda_C=\hbar/(mc)$ imposes a fundamental limitation on its position, that we can call a {\it quantum of length}\/, denoted $Q_1$. This $Q_1$, which is particle--dependent, is of a fundamentally different nature than the {\it geometric}\/ quantum of length $L_P$. On configuration space $\mathbb{R}^D$, this gives rise to a quantum $Q_{D-1}$ of $(D-1)$--dimensional volume within the leaf $\mathbb{L}_n$, with measure (proportional to) $\lambda_C^{D-1}$, and to a quantum of length $Q_1$ along the normal coordinate. 

In the presence of more than one particle species with different masses, each mass $m_i$ defines one value of the quantum $Q_{D-1}^{(i)}$. Then a  quantum of volume that remains valid for all particles is the largest value of all those  $Q_{D-1}^{(i)}$. This is the quantum of volume determined by the lightest particle.

Let us now elucidate how quantum states $\psi$ can arise as equivalence classes of different classical trajectories. By Eq. (\ref{fakto}) we have to account for the appearence of the normal wavefunction $\psi_n$ and of the tangential wavefunction $\psi_t$. 

Starting with $\psi_t$, let us consider all the different classical trajectories traversing any one quantum of volume $Q_{D-1}$ within a leaf $\mathbb{L}_n$. The allowed values of the momentum carried by those trajectories are those compatible with the uncertainty principle. Since the particle has been spatially localised to an accuracy of $\lambda_C$ along each tangential coordinate, the corresponding momentum can be specified to an accuracy of $\hbar/\lambda_C$. Therefore, corresponding to a spatial quantum of volume $Q_{D-1}$ in the leaf, we have a quantum of volume $P_{D-1}=(\hbar/\lambda_C)^{D-1}$ in momentum space. 

We are now in a position to state a postulate:

{\it All the different classical trajectories traversing any quantum of volume $Q_{D-1}$ in the leaf\/ $\mathbb{L}_n$, and simultaneously traversing a quantum $P_{D-1}$ in tangential momentum space, are to be regarded as different representatives of just one tangential  state $\psi_t$}\/. 

An analogous postulate for the normal coordinate reads:

{\it All classical trajectories traversing any quantum of length $Q_1$ along the normal coordinate $n$, and simultaneously traversing the corresponding quantum $P_1$ in normal momentum space, make up one normal state $\psi_n$}\/. 

In support of the above postulate, let us return to Eq. (\ref{diktionaer}), where the mechanical combination $m\omega/\hbar$ has been identified with the thermodynamical quotient $s/(2k_B)$. The constant $s$, defined in Eq. (\ref{teilo}), carries the dimensions of entropy $\times$ $x^{-2}$, so $s/(2k_B)$ has the dimensions $x^{-2}$. Thus $s/(2k_B)$ is homogeneous to the inverse square of the Compton wavelength, $\lambda_C^{-2}$. 

On the other hand, the constant $s$ (and the frequency $\gamma$ in (\ref{diktionaer})) are all the data one needs in order to univocally specify the irreversible thermodynamics that is dual to the given quantum mechanics. The previous statement, which holds exactly true in the harmonic approximation of section \ref{dos}, is raised to the category of a principle in the above postulate. Indeed, let us assume going beyond the harmonic approximation in mechanics. In the thermodynamical dual theory, this is equivalent to considering terms beyond quadratic in the Taylor expansion (\ref{teilo}). Higher derivatives ${\rm d}^3S/{\rm d}x^3$, ${\rm d}^4S/{\rm d}x^4$, etc, evaluated at the equilibrium point, simply introduce new constants $s_3$, $s_4$, etc, which can be dimensionally accounted for in terms of just two physical constants, namely $k_B$ and $\lambda_C$. Up to a set of {\it dimensionless}\/ coefficients, all the data we need in the irreversible thermodynamics can be constructed in terms of $k_B$ and powers of $\lambda_C$. 

These arguments render our above postulate a very plausible statement. Moreover, they provide an estimate of the entropy increase ({\it i.e.}\/, of the amount of information loss) involved in the lumping together of many classical trajectories into just one quantum state. Namely,  {\it the increase in entropy $\Delta S$ due to the formation of one equivalence class of classical trajectories is a positive multiple of $\lambda_C^2$ times the coefficient $s$}\/,
\begin{equation}
\Delta S=ns\lambda_C^2,\qquad n>0,
\label{jof}
\end{equation}
{\it where $n$ is a  dimensionless number}\/. (Admittedly, our arguments leave $n$ undetermined, although one could resort to Landauer's principle \cite{LANDAUER} in order to argue that $n$ must be of order unity). More importantly, the surface density of entropy $s$ can be naturally identified, via Eq. (\ref{jof}), with the entropy increase $\Delta S$ due to the formation of quantum states as equivalence classes \cite{THOOFT1, THOOFT2}. In other words, {\it the dissipation that is inherent to irreversible thermodynamics has a natural counterpart in quantum mechanics}\/.

Having described the dissipative mechanism whereby classical trajectories organise into quantum states, we go next to a counting of the number of quantum states. Since the leaf $\mathbb{L}_n$ has been assumed compact, it encloses a finite number $N_n$ of volume quanta $Q_{D-1}$. Tentatively identifying this number $N_n$ with the (complex) dimension of the tangential Hilbert space ${\cal H}_t$, we immediately realise that the quantum of momentum $P_{D-1}$ is contained an infinite number of times within tangential momentum space (this is however a {\it countable}\/ number of times). Indeed the momenta may grow to arbitrarily large values. Therefore, the tangential Hilbert space ${\cal H}_t$ is infinite--dimensional, and separable.

On the other hand, the dimension of the normal Hilbert space ${\cal H}_n$ is infinite already from the start (again a countable infinity, hence 
${\cal H}_n$ is separable). The reason for this is the noncompactness of  $\mathbb{R}^D$: the normal coordinate $n$ must cover an interval of infinite length.\footnote{In case more than just one normal coordinate is needed, this statement is to be understood as meaning the sum of all the lengths so obtained.} This implies that the normal coordinate encloses an infinite (though countable) number of length quanta $Q_1$. Multiplication by the number of independent momentum quanta $P_1$ does not alter this separable, infinite--dimensionality of ${\cal H}_n$.

Altogether, the complete Hilbert space ${\cal H}$ of quantum states is the tensor product ${\cal H}_t\otimes{\cal H}_n$. However, because it singles out the normal coordinate $n$, one might worry that our construction depends on the particular choice of a leaf $\mathbb{L}_n$ within the foliation. Now the only possible difference between any two leaves $\mathbb{L}_{n_1}$ and $\mathbb{L}_{n_2}$ is the value of their $(D-1)$--dimensional volume. Hence the numbers of volume quanta $N_{n_1}$ and $N_{n_2}$ they enclose may be different---but they are both finite. This possible difference is washed away upon multiplication by the (countably infinite) number of momentum quanta $P_{D-1}$ corresponding to each leaf. The dimension of ${\cal H}_t$ is therefore countably infinite regardless of the point, $n_1$ or $n_2$, along the radial coordinate---that is, regardless of which leaf is considered.\footnote{We should remark that the assumption of compactness of the leaves $\mathbb{L}_n$ can be lifted without altering our conclusions. A noncompact leaf encloses an infinite (yet countable) number of volume quanta $Q_{D-1}$. Upon multiplication by an infinite (yet countable) number of momentum--space quanta $P_{D-1}$, the dimension of the tangent Hilbert space ${\cal H}_t$ remains denumerably infinite. This form of holography in which the leaves are noncompact replaces the notion of {\it inside vs. outside}\/ the leaf with the equivalent notion of {\it one side of the leaf vs. the other side}\/. One should not dismiss this possibility as unphysical: the constant potential, for example, can be regarded as having either compact or noncompact equipotential submanifolds.}

As explained in ref. \cite{ACOSTA}, determining the tangential wavefunctions $\psi_t$ does not require a knowledge of the specific dynamics under consideration. Instead, this tangential dependence is univocally fixed by the geometry of the leaves $\mathbb{L}_n$. In more technical terms, the wavefunctions $\psi_t$ must provide a complete orthonormal set for a unitary, irreducible representation of the isometry group of the leaves $\mathbb{L}_n$. Moreover, as argued in ref. \cite{ACOSTA}, the modulus squared $\vert\psi\vert^2$, evaluated at the value $n$, is proportional to the surface density of entropy flux across the leaf $\mathbb{L}_n$.

\section{Quantum uncertainty vs. the second law}\label{cuatro}

Just as Planck's constant $\hbar$ represents a coarse--graining of phase space into cells of minimal volume, or quanta of action, so does Boltzmann's constant $k_B$ represent a {\it quantum of entropy}\/. This implies that any process must satisfy the condition
\begin{equation}
\Delta S=Nk_B, \qquad N\in\mathbb{N}.
\label{volman}
\end{equation}
The above expresses a quantised form of the second law of thermodynamics. The extreme smallness of the numerical value of $k_B$ in macroscopic units makes this quantisation macroscopically unobservable. In particular, unless $N=0$, the second law becomes
\begin{equation}
\Delta S\geq k_B.
\label{bolman}
\end{equation}
In this form, the second law is actually a rewriting of the quantum--mechanical uncertainty principle for the canonical pair $E, t$:
\begin{equation}
\Delta E\Delta t\geq \frac{\hbar}{2}.
\label{uncer}
\end{equation}
Of course, this derivaton of the uncertainty relation $\Delta E\Delta t\geq \hbar/2$ is heuristic, because time is a parameter in quantum mechanics.
It is only in the limit $k_B\to 0$ that the second law (\ref{bolman}) reduces to its classical formulation $\Delta S\geq 0$. The limit $k_B\to 0$ is the thermodynamical counterpart of the usual semiclassical limit $\hbar\to 0$ of quantum mechanics. 

We conclude that the equivalence between Eqs. (\ref{bolman}) and (\ref{uncer}) is a consequence of our basic postulate (\ref{refinatissimo}). In other words, the second law (\ref{bolman}) expresses, in the thermodynamical theory, the same statement as the uncertainty principle (\ref{uncer}) expresses in the quantum--mechanical theory. 

Our correspondence implies that, while one needs two canonical variables $E,t$ in order to express the uncertainty principle in the quantum theory, just one variable $S$ is needed in order to write the second law. An equivalent way of saying this is that {\it entropy is a selfconjugate variable}\/: one does not have to multiply it with a canonical variable (say, $\xi$) in order to obtain a product $\xi S$ carrying the dimensions of the quantum $k_B$. The variable $S$ already carries the dimensions of its corresponding quantum $k_B$.

\section{Discussion}\label{cinco}

The holographic principle of quantum gravity states that there fits at most 1 bit of information into each quantum of area $L_P^2$ in configuration space, where $L_P$ is Planck's length. For quantum mechanics, in section \ref{tres} we have postulated that

{\it There fits at most 1 quantum state into each quantum of volume $(\lambda_C)^{2D}$ in phase space, whereby the Compton length $\lambda_C$ of the particle in question extends once along each coordinate $q$ and once along each conjugate momentum $p$ in a $2D$--dimensional phase space}\/.

Thus our postulate is conceptually analogous to the holographic principle of quantum gravity. We should stress, however, that our postulate does not follow from, nor does it imply, the holographic principle of quantum gravity.

We can summarise our construction as follows. Let a quantum--mechanical system be given in configuration space $\mathbb{R}^D$. Let this latter space be foliated as per $\cup_{n}\mathbb{L}_n$, where each leaf $\mathbb{L}_n$ is an equipotential submanifold, in dimension $D-1$, of the given mechanical potential function $V(q_i)$. Assume that each leaf $\mathbb{L}_n$ encloses a finite $D$--dimensional volume $\mathbb{V}_n$, so $\partial\mathbb{V}_n=\mathbb{L}_n$. Then quantum states in $\mathbb{V}_n$ are equivalence classes of different classical trajectories. These equivalence classes comprise all those classical trajectories that fit into one given quantum of volume in configuration space, with the corresponding momenta inside the corresponding quantum in momentum space. No quantum particle can be located to an accuracy better than its Compton wavelength.\footnote{Unless, of course, one is willing to allow for pair creation out of the vacuum, thus quitting quantum mechanics and entering field theory.} Hence a physically reasonable unit for defining this quantum of length (and thus areas and volumes) is the Compton wavelength. Configuration space is subdivided into many such elementary volume quanta, each one of them (with the corresponding quanta in momentum space) defining one different quantum state.

The quantisation of phase--space area by Planck's constant $\hbar$ proceeds along lines that are somewhat similar to ours, although not exactly identical. We recall that, semiclassically, the (symplectic) area element ${\rm d} p\wedge{\rm d}q$, divided by $\hbar$, gives the number of different quantum states fitting into that area element. However, the coordinate width ${\rm d}q$ may be arbitrarily squeezed, provided the momentum ${\rm d} p$ is correspondingly enlarged, and viceversa. 

On the contrary, our construction makes use of the Compton wavelength $\lambda_C$ as a fundamental quantum of length (for the specific particle considered), below which no sharper localisation is possible: there is no squeezing the particle below this lower limit. This gives rise to an arrangement of different classical trajectories into equivalence classes that, following ref. \cite{THOOFT2}, we identify with quantum states. This is an irreversible, dissipative mechanism that exhibits the emergent nature of quantum mechanics. The Hilbert space of quantum states is 
determined as described in section \ref {tres}.

Under our correspondence, an irreversible thermodynamics can be mapped into a quantum mechanics, and viceversa. This correspondence may be regarded as {\it dictionary}\/ that allows one to switch back and forth between a {\it quantum--mechanical picture}\/ and a {\it thermodynamical picture}\/ of one and the same physics. 

A key point to remark is the following. Thermodynamical approaches to quantum theory are well known \cite{DEBROGLIE, MATONE}. In particular, the link between (complex--time) quantum mechanics, on the one hand, and the {\it equilibrium}\/ statistical mechanics of the Gibbs ensemble, on the other, has been known for long. We should stress that we have {\it not}\/ dwelled on this long--established connection. Rather, the new correspondence explored here is that between (complex--time) quantum mechanics, and the {\it classical}\/ thermodynamics of {\it irreversible}\/ processes. {\it Classicality}\/ of the thermodynamics means that $\hbar$ does not appear on the thermodynamical side of the correspondence, its role being played instead by Boltzmann's constant $k_B$. {\it Irreversibility}\/ implies the existence of dissipation, as befits the presence of quantum effects.

\end{document}